\documentclass[prb,twocolumn,floatfix,superscriptaddress]{revtex4-1}
\usepackage[pdftex,plainpages=false,colorlinks=true,linkcolor=blue, citecolor=blue, urlcolor=blue]{hyperref}
\usepackage{amsfonts}
\usepackage{amsmath}
\usepackage{amssymb}
\usepackage{graphicx}
\usepackage{natbib}
\usepackage{color}
\usepackage{placeins}

\begin{document}
 
\title{Signatures of the Mott transition in the antiferromagnetic state of the two-dimensional Hubbard model}
\author{L. Fratino}
\affiliation{Department of Physics, Royal Holloway, University of London, Egham, Surrey, UK, TW20 0EX}
\author{P. S\'emon}
\affiliation{D\'epartement de physique and Regroupement qu\'eb\'equois sur les mat\'eriaux de pointe, Universit\'e de Sherbrooke, Sherbrooke, Qu\'ebec, Canada J1K 2R1}
\affiliation{Computational Science Initiative, Brookhaven National Laboratory, Upton, NY 11973-5000, USA}
\author{M. Charlebois}
\affiliation{D\'epartement de physique and Regroupement qu\'eb\'equois sur les mat\'eriaux de pointe, Universit\'e de Sherbrooke, Sherbrooke, Qu\'ebec, Canada J1K 2R1}
\author{G. Sordi}
\affiliation{Department of Physics, Royal Holloway, University of London, Egham, Surrey, UK, TW20 0EX}
\author{A.-M. S. Tremblay}
\affiliation{D\'epartement de physique and Regroupement qu\'eb\'equois sur les mat\'eriaux de pointe, Universit\'e de Sherbrooke, Sherbrooke, Qu\'ebec, Canada J1K 2R1}
\affiliation{Canadian Institute for Advanced Research, Toronto, Ontario, Canada, M5G 1Z8}

\date{\today}

\begin{abstract}
The properties of a phase with large correlation length can be strongly influenced by the underlying normal phase. 
We illustrate this by studying the half-filled two-dimensional Hubbard model using cellular dynamical mean-field theory with continuous-time quantum Monte Carlo. Sharp crossovers in the mechanism that favors antiferromagnetic correlations and in the corresponding local density of states are observed. These crossovers occur at values of the interaction strength $U$ and temperature $T$ that are controlled by the underlying normal-state Mott transition. 
\end{abstract}
 
\maketitle

\section{Introduction}

A striking manifestations of strong interactions in quantum materials is the Mott transition~\cite{mott, ift, Anderson:1987}, a first-order transition between a half-filled band metal and an insulator. The Mott transition does not break spatial symmetries, but it is often masked by, or is found in close proximity to, broken symmetry states, notably the N\'eel antiferromagnetic  state (AF)~\cite{mott, ift}. Indeed, in the Mott insulator the electrons are localized, so local moments tend to order magnetically at low temperature via the superexchange mechanism~\cite{AndersonSE, FazekasBook:1999}. Experimentally, the Mott transition accompanied by AF occurs in materials with partially filled narrow orbitals, both with a three-dimensional (3D)  structure, such as V$_2$O$_3$~\cite{mcwhan, ift, limelette}, CsC$_{60}$~\cite{TakabayashiScience2009, Ihara:2010, Alloul_arxiv2016}, and with a quasi-two-dimensional (2D) layered structure, such as superconducting copper oxides~\cite{Bednorz:1986, Anderson:1987, keimerRev} and organics~\cite{Lefebvre:2000, LimeletteBEDT:2003, kagawaPRB, kagawa1, Gati2016}. Ultracold atoms in optical lattices~\cite{ColdAtomsRMP, Esslinger:2010} also offer a platform to study the Mott transition~\cite{Jordens:2008, Schneider:2008, Hofrichter:PRX2016, Cocchi:PRL2016} and, recently, its interplay with AF correlations~\cite{Greif:2013, Hart:2015, Parsons:2016, Boll:2016, Cheuk:2016}. 

The Hubbard Hamiltonian, describing the competition between nearest-neighbor hopping $t$ and on-site screened electron-electron interaction $U$, is the simplest model that captures the Mott transition and its interplay with AF. It is known~\cite{AMJulich, AntoineLesHouches} that the half-filled model in 3D shows a single AF phase, namely at $T=0$ one does not encounter a phase transition as the ratio $U/t$ is increased. Nevertheless, the mechanism that makes the normal state unstable to AF as $T$ decreases is described differently in the two limits: at small $U/t$, AF arises from cooling a metal, and thus stems from nesting of the Fermi surface, whereas for large $U/t$, AF originates from cooling a Mott insulator, and thus stems from superexchange between localized spins~\cite{AMJulich, AntoineLesHouches}. These mechanisms are referred to as {\it Slater} and {\it Heisenberg} respectively. In the Slater regime, increasing $U/t$ leads to an increase in the N\'eel temperature $T_N$, whereas in the Heisenberg regime increasing $U/t$ leads to a decrease in $T_N$. This is one of many {\it qualitative} differences between AF at small and large $U/t$.

Above $T_N$, depending on the strength of lattice or hopping-induced frustration, the Mott transition can be either apparent or hidden by the N\'eel state. When the Mott transition is hidden, one is left with a crossover from metallic to insulating state as $U/t$ is increased. In both cases one expects that the properties of the AF depend on the normal state from which it emerges~\cite{rmp}. 

The above results for the basic experimental phenomenology of 3D systems as well as predictions (e.g.~\cite{rmp, kotliarRMP, WernerAdiabCooling2005, TarantoPRB2012}) have been obtained from the dynamical mean-field theory (DMFT)~\cite{rmp} solution of the Hubbard model. However, the strong momentum dependence of the self-energy in 2D makes DMFT inadequate in that case. Cluster extensions of DMFT are a way to include some of this momentum dependence. These methods and others have been used to study the AF phase~\cite{Hirsch:1985, White:1989, lkAF, Paiva:2001, Dupuis2004, maier, kyung, Paiva2010, Shafer2015, Sato2016}, but the influence of the normal-state Mott transition, if any, on  the AF phase has been less investigated~\cite{Tocchio2016} and remains a challenge. Here, we contribute to decipher the interplay between Mott transition and AF by studying the finite temperature aspects of both normal and AF states of the half-filled 2D Hubbard model using a cluster extension of DMFT~\cite{maier, kotliarRMP, tremblayR}. Our calculations reveal crisp differences between weakly and strongly interacting AF that are linked to the normal state Mott transition hidden beneath the AF dome. Remarkably, the Mott transition controls the sharp crossover that we observe between a potential-energy driven AF at small $U$, and kinetic-energy driven AF at large $U$. This question of the origin of the stability of the AF state relative to the normal state has hitherto received little attention compared with the same question for superconductivity~\cite{Chester:1956, LeggettCondEn, scalapinoCondEn, Norman:2000, Molegraaf2002, deutscher2005, andersonBOOK, Hirsch1, Demler1998, maierENERGY, carbone2006, kyungPAIRING, millisENERGY, ssht, LorenzoSC}. Yet, both questions are related to the role of the normal-state Mott transition. 

Note that in 2D, thermal fluctuations preclude long-range order so that the AF transition occurs at $T_N=0$, in accordance with the Mermin-Wagner-Hohenberg theorem~\cite{MWtheorem,Hohenberg:1967}. Nevertheless the AF correlation length $\xi$ {\it does} begins to grow exponentially at an onset temperature that can be quite sharply defined. In the cluster extension of DMFT that we use, the onset of exponential growth of $\xi$ is replaced by long-range order at a temperature that we call $T_N^d$, where the $d$ reminds us that this is the {\it dynamical} mean-field temperature. 
In Sec.~\ref{Sec:Benchmarks} we provide benchmarks showing that the mostly local physical properties that we study in the ordered state are close to those one would obtain for large systems with large $\xi$. 

We begin by a short description of the model and method in Sec.~\ref{Sec:ModelMethod} (that is further detailed in Appendix \ref{Sec:Method}) and we then set the stage in Sec.~\ref{Sec:PhaseDiagram} with the phase diagram. The main results are in Secs.~\ref{energetics} and \ref{Sec:DOS_P} that discuss the crossovers in the antiferromagnetic state of, respectively, the energetics and of the density of states. In Sec.~\ref{concl} we present the conclusions. Appendix \ref{Sec:mz} contains further details on the order parameter, appendix \ref{Sec:Energy} on the calculation of the energy, and appendix \ref{Sec:DOS} on results for the density of states.


\section{Model and method}\label{Sec:ModelMethod} 
We consider the single-band Hubbard model on the square lattice in 2D, $H=-\sum_{ij\sigma}t_{ij}c_{i\sigma}^\dagger c_{j\sigma}  +U\sum_{i} n_{i\uparrow } n_{i\downarrow } -\mu\sum_{i\sigma} n_{i\sigma}$, where $c^{\dagger}_{i\sigma}$ ($c_{i\sigma}$) create (destroy) an electron of spin $\sigma$ on site $i$, $n_{i\sigma}=c^{\dagger}_{i\sigma}c_{i\sigma}$ is the number operator, $\mu$ is the chemical potential and $U$ the onsite Coulomb repulsion. We take $t_{ij}=t=1$ for nearest neighbor hopping. We aim to study the local quantum fluctuation induced by $U$ on the same footing as the short-range correlations, hence we solve this model using cellular dynamical mean-field theory (CDMFT)~\cite{maier, kotliarRMP, tremblayR}. CDMFT takes a cluster, here a 2$\times$2 plaquette, out of the lattice and replaces the missing lattice environment by a self-consistent bath of noninteracting electrons. We solve the impurity (plaquette in a bath) problem using the statistically exact continuous time quantum Monte Carlo method~\cite{millisRMP} based on the hybridization expansion of the impurity action, best suited for the large values of $U$ and low $T$ that are mostly considered here. 

We consider normal and AF states. In the latter phase, symmetry breaking is allowed only in the bath and not in the cluster. We accelerate the calculation by taking advantage of $C_{2v}$ group symmetry with mirrors along the diagonals of the plaquette~\cite{patrickCritical, patrickERG} (see also Appendix~\ref{Sec:Method}). We consider only the half-filled model.
\begin{figure*}[ht!]
\centering{
\includegraphics[width=1\linewidth, clip=]{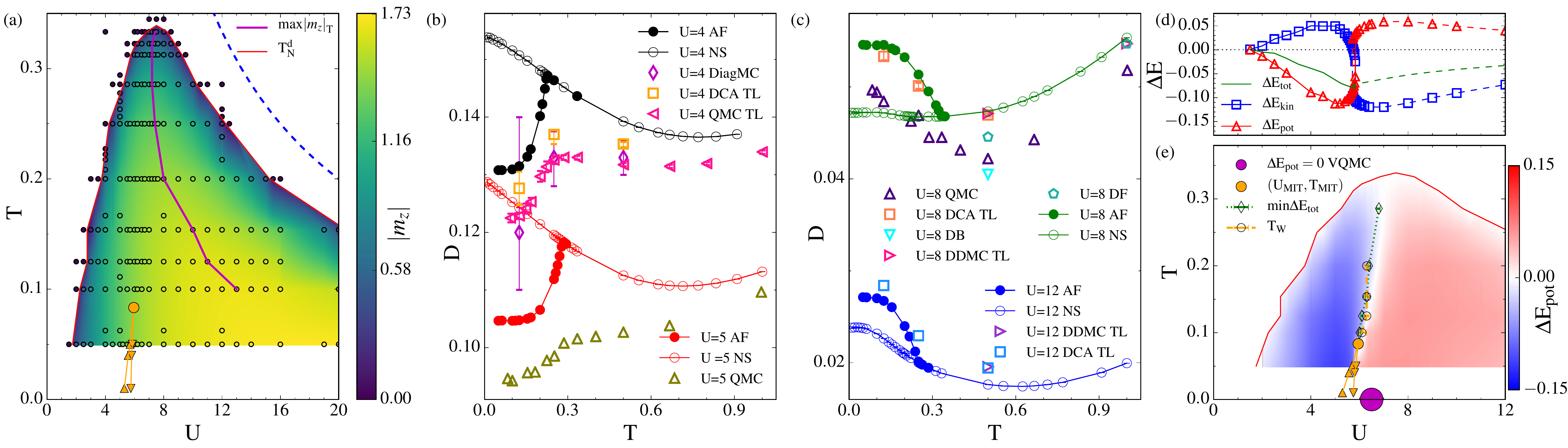}}
\caption{(a) N\'eel temperature $T^d_N$ versus $U$ at $n=1$. Color corresponds to the magnitude of the staggered magnetisation $m_z$ (raw data is in Appendix~\ref{Sec:mz}). Magenta line indicates the maximum of $m_z$ at fixed $T$. The AF phase is defined by the loci where $m_z \neq 0$ and is delimited by $T_N^d$. At low temperature the normal state shows a first-order transition between a metal and a Mott insulator (orange triangles lines) that ends at a critical endpoint at $(U_{\rm MIT},T_{\rm MIT})$ (orange filled circle). (b, c) Double occupancy $D$ versus $T$ at $U=4, 5 <U_{\rm MIT}$ and $U=8, 12 > U_{\rm MIT}$, for both AF and normal states (filled and open circles, respectively). For benchmarks, we show data from alternative methods: diagrammatic Monte Carlo from Ref.~\onlinecite{LeBlancPRX} (diamonds), DCA extrapolated to infinite lattice from Ref.~\onlinecite{LeBlancPRX} (squares), determinantal QMC on $10^2$ lattice from Ref.~\onlinecite{ Paiva2010} (up triangles) and extrapolated to thermodynamic limit~\cite{Paiva:2001} (left triangles), dual boson scheme from Ref.~\onlinecite{vanLoonDB} (down triangle), diagrammatic determinant Monte Carlo extrapolated to thermodynamic limit from Ref.~\onlinecite{janDMC2016} (right triangles) and dual fermion scheme from Ref.~\onlinecite{janDMC2016} (pentagons). (d) Difference in potential, kinetic and total energies (red, blue and green lines) between  the AF and normal state versus $U$ at $T=1/20$. (e) Phase diagram $T-U$ with colormap of $\Delta E_{\rm pot}$. A change of sign in $\Delta E_{\rm pot}$ occurs along a line connecting $U_{\rm MIT}$ and $T_N^{d, \rm max}$. It accompanies the loci of the largest condensation energy (green diamonds), which in turn correlates with the normal state Widom line~\cite{water1, ssht, sshtRHO}, emanating from the Mott endpoint, and determined by max~$dD/dU|_T$ (open circles). See also Fig.~\ref{figSM6} in Appendix~\ref{Sec:Energy}. For benchmark, we show with a magenta circle the $T=0$ variational QMC calculation of Ref.~\onlinecite{Tocchio2016} for $\Delta E_{\rm pot}=0$. 
}
\label{fig1}
\end{figure*}
%

\section{Phase diagram}\label{Sec:PhaseDiagram} 
Figure~\ref{fig1}a shows $T_N^d$ versus interaction strength $U$. The magnitude of the staggered magnetisation $m_z$ is color-coded. $T_N^d(U)$ has a dome shape.  Similarly to the 3D case, this behavior alone already suggests that although there is a single AF phase, there are qualitative differences as a function of $U$: the initial rise of $T_N^d$ occurs because of nesting of the Fermi surface, whereas at large $U$ superexchange $J=4t^2/U$ leads to the decrease of $T_N^d$ with $U$~\cite{AMJulich, Dupuis2004}. Our results for $T_N^d$ are close to the onset of exponential behavior~\cite{White:1989,Vilk:1997, Dupuis2004, trilex1, trilex2} of $\xi$, i.e. close to the crossover to the renormalized-classical regime. The line connecting the maximum of $m_z$ at fixed temperatures (magenta line) approximately follows $J$ (blue dashed line), thereby indicating that superexchange drives the N\'eel ordering at large $U$.  As $T\rightarrow 0$, $m_z$ saturates to its maximum value at large $U$. Key indicator of the difference between weakly and strongly interacting AF is the nature of the normal state just above $T_N^d$.  At small $U$ the N\'eel state arises from cooling a metal, and at large $U$ it arises from cooling a Mott insulator~\cite{phk}. The normal state metal-insulator {\it crossover} above $T_N^d$ is controlled by the Mott metal-insulator {\it transition} (MIT) that would occur at low $T$ in the normal state if AF was not permitted. That transition has first-order character and ends in a critical endpoint at $(U_{\rm MIT}, T_{\rm MIT})$: orange triangles in Fig.\ref{fig1}a indicate the coexistence region between a metal and a Mott insulator, as obtained by isothermal hysteresis loops in the double occupancy as a function of $U$~\cite{phk}. 
We show below that even though this Mott transition is hidden beneath $T_N^d$, it controls the striking differences in the energetics and in the density of states (DOS) between AF at weak and strong interaction.  

\section{Some benchmarks}\label{Sec:Benchmarks}
Double occupancy $D=\langle n_{i\uparrow} n_{i \downarrow} \rangle$ measures the degree of electronic correlations. It is shown in Fig.~\ref{fig1}b,c as a function of $T$ for values of $U$ below and above $U_{\rm MIT}$, in both the AF and normal states (filled and open circles, respectively). The normal state is unstable but, as in any mean-field theory, it can be continued for $T<T_N^d$ by simply forcing $m_z=0$. As benchmarks, we also plot the values of $D$ obtained from alternative approaches~\cite{LeBlancPRX, vanLoonDB, Paiva:2001, Paiva2010, janDMC2016}. For $T\ll T_N^d$, AF results are closer to these benchmarks than normal state results. This is expected because, as far as local quantities are concerned, a state with a finite but exponentially large $\xi$ is closer to an AF state than a normal state with $\xi$ at most one lattice spacing. The kink in our $D(T)$ at $T_N^d$ is replaced in the benchmarks by a shallow crossover. These benchmarks confirm the validity of our CDMFT approach, which, for local quantities such as $D$, converges exponentially fast with cluster size~\cite{BiroliExp2005, kotliarRMP}. As expected, the agreement improves with increasing $U$, since states are then more localized, and it also improves with increasing temperature above $T_N^d$ where $\xi$ is smaller. 

\section{Energetics}\label{energetics} 
In the normal state, $D(T)$ shows a minimum. This occurs because increasing the local moment (decreasing $D$) increases spin entropy upon heating from $T=0$, while the non-interacting limit must be reached at very large $T$~\cite{antoine, rmp}. As $T$ is reduced below $T_N^d$, the normal state becomes unstable to AF. We find a sharp difference between weak and strong interactions. For $U<U_{\rm MIT}$ (Fig.~\ref{fig1}b), the N\'eel state not only suppresses $D$ compared with $D$ in the normal state, it also reverses the slope of $D(T)$ (i.e. for $T<T_N^d$: $dD/dT>0$ in the AF state and $dD/dT<0$ in the normal state). On the other hand,  for $U>U_{\rm MIT}$, (Fig.~\ref{fig1}c), $D$ is increased in the AF state. 

The above contrasting results show that the AF state leads to a potential energy $E_{\rm pot}=UD$ decrease when $U<U_{\rm MIT}$ and to a potential-energy increase when $U>U_{\rm MIT}$. This goes at the heart of the origin of the $T=0$ stability of the AF state relative to the normal state. At low $U$, the energetics agree with the expectation of a Slater insulator and static mean-field theory, where the order parameter $m_z$ corresponds to a larger local moment or decrease in $D$. In contrast, at large $U$ in the Heisenberg limit, the kinetic energy is minus twice the potential energy~\cite{FazekasBook:1999}, and thus the AF state is stabilized by a kinetic-energy gain. This is illustrated in Fig.~\ref{fig1}d where the difference in potential, kinetic and total  energies between the AF and the normal state, is plotted versus $U$ for $T<T_{\rm MIT}$ (see also Appendix~\ref{Sec:Energy}). Crucially, the critical $U$ at which $\Delta E_{\rm pot}$  and $\Delta E_{\rm kin}$ cross zero, and $\Delta E_{\rm tot}$ is largest, is determined by the Mott transition. This is one of our main findings. 

An even more remarkable finding is apparent from Fig.\ref{fig1}e, where $\Delta E_{\rm pot}$ is color coded for the AF region: the change of sign in $\Delta E_{\rm pot}$ (see sharp white region) that signals the crossover from weak to strong interactions occurs at the normal state Mott transition for $T<T_{\rm MIT}$, and continues for $T>T_{\rm MIT}$ in a nontrivial crossover connecting the Mott endpoint to approximately $T_N^{d, \rm max}$. The region where $\Delta E_{\rm pot}(T)$ crosses zero accompanies the loci of the largest condensation energy (green diamonds), which in turn correlate with the normal state Widom line~\cite{water1, ssht, sshtRHO} emanating out of the endpoint (as determined by max~$dD/dU|_T$, and indicated by open circles). Recent variational QMC calculations~\cite{Tocchio2016} find that $\Delta E_{\rm pot}$ crosses zero between $U=6$ and $U=7$ (see magenta circle at $T=0$ in Fig.~\ref{fig1}e). This benchmark supports our results. A similar crossover from potential-energy driven to kinetic-energy driven mechanisms is also observed in studies of BCS-BEC crossover in the attractive Hubbard model~\cite{AMJulich, kyungPAIRING, GargBCS2005, ToschiBCS2005, TarantoPRB2012}.

\begin{figure}[t]
\centering{
\includegraphics[width=0.98\linewidth]{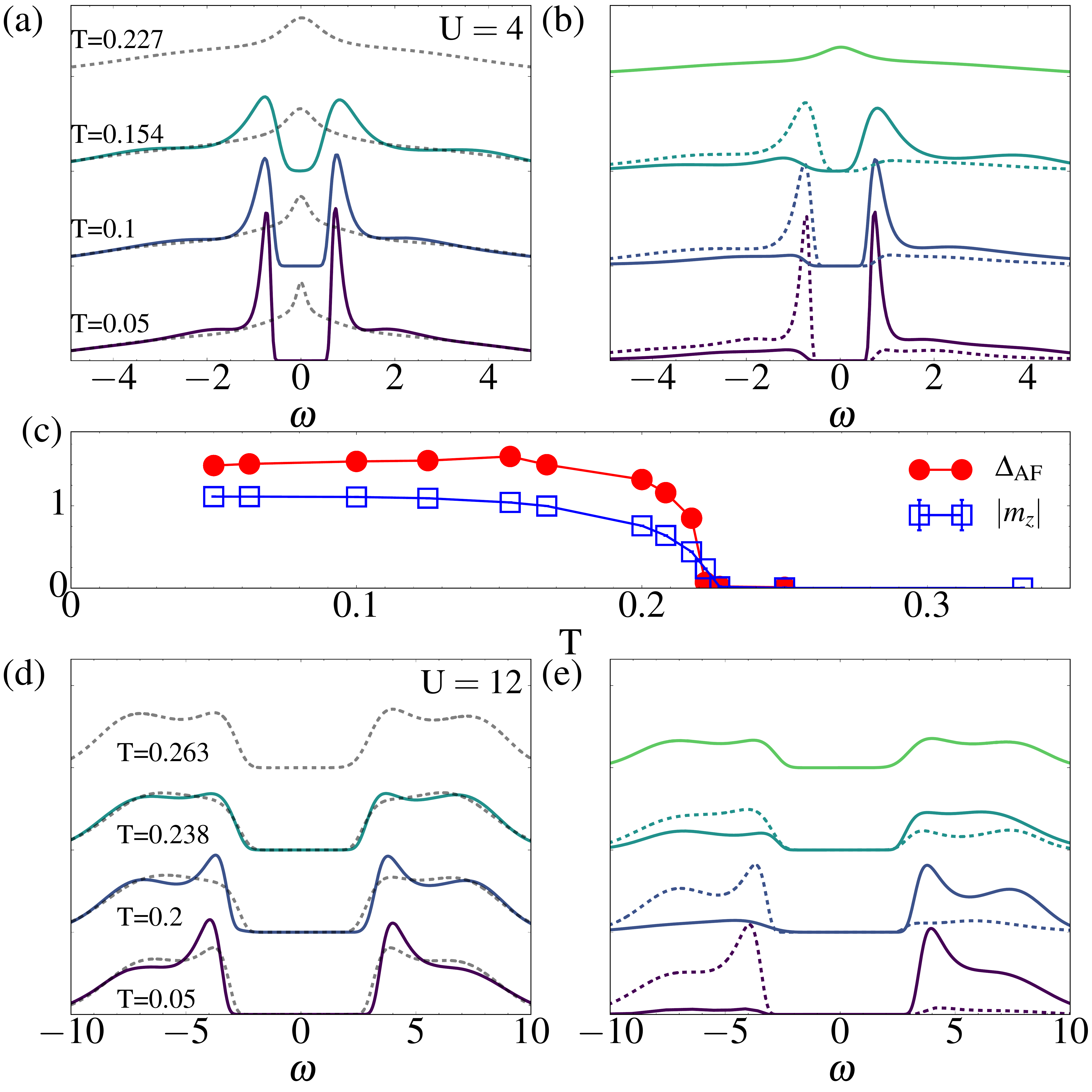}}
\caption{Temperature evolution of (a) $N(\omega)$ and (b) its spin projections for $U=4 < U_{\rm MIT}$, as obtained from analytically continued data~\cite{DominicMEM}. AF (normal) state DOS are shown with color (dashed grey) lines. (c) AF gap $\Delta_{\rm AF}$ versus $T$ for $U=4$, as measured by the distance between the Bogoliubov peaks. The closure of $\Delta_{\rm AF}$ follows the decrease of $m_z$. (d,e) same as (a,b) but for $U=12>U_{\rm MIT}$. 
}
\label{fig2}
\end{figure}

\section{Density of states}\label{Sec:DOS_P}
\begin{figure}[t]
\centering{
\includegraphics[width=0.995\linewidth]{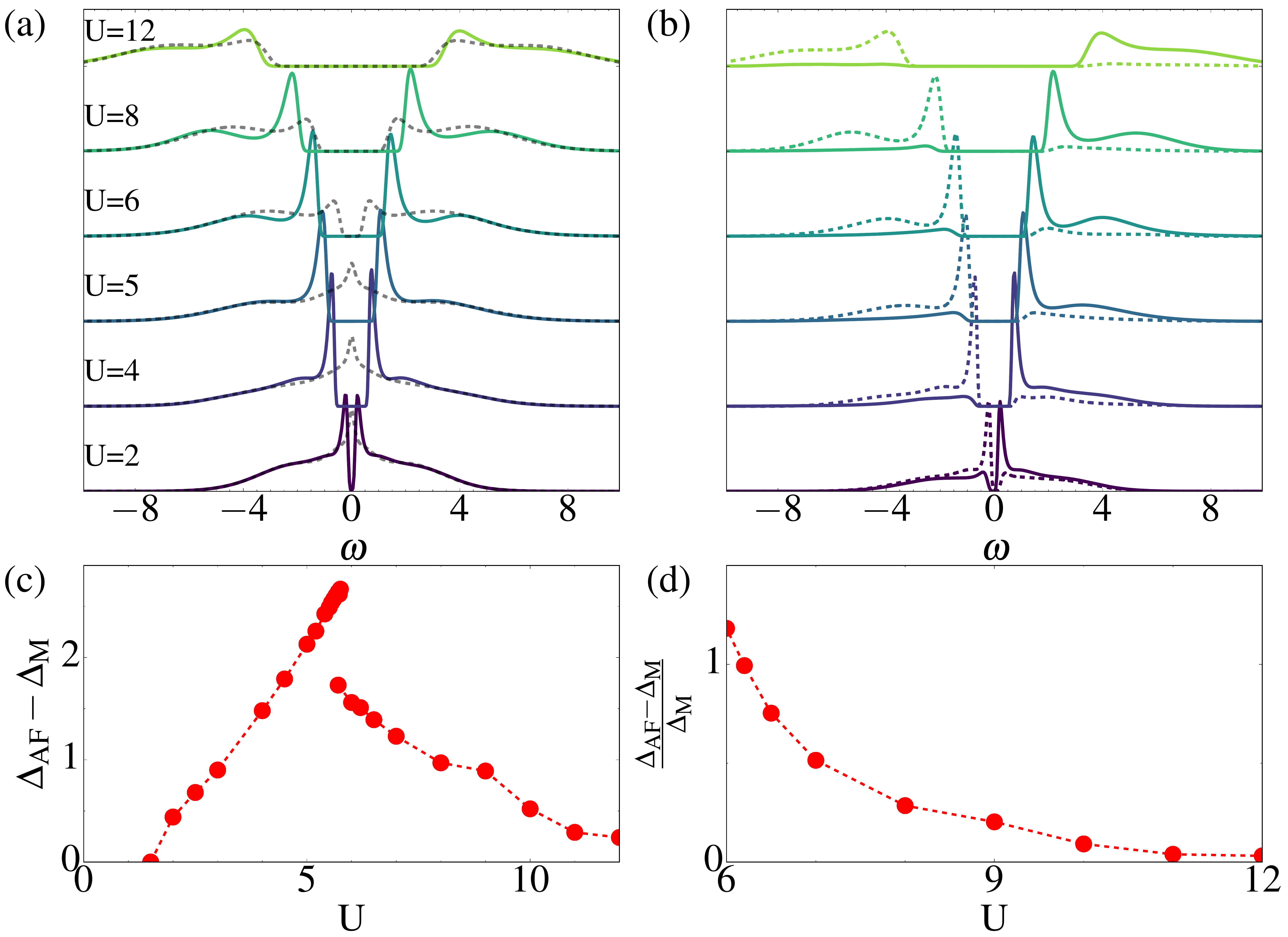}}
\caption{(a) $N(\omega)$ in the AF state along with (b) its two spin projections, for $T=1/20$ and different values of $U$. Normal state solutions are shown with grey lines. See also Fig.~\ref{figS7} in Appendix~\ref{Sec:DOS}. (c) Difference between the AF gap $\Delta_{\rm AF}$ and the normal state Mott gap $\Delta_{\rm M}$, measured by the distance between the two Hubbard bands, versus $U$ for $T=1/20$. (d) $(\Delta_{\rm AF}-\Delta_{\rm M})/\Delta_{\rm M}$ versus $U$ for $T=1/20$. 
}
\label{fig3}
\end{figure}

The underlying normal state Mott transition also leads to qualitative differences in the AF state between the local DOS $N(\omega)$ observed at weak and strong interaction. This is illustrated in Figs.~\ref{fig2} and~\ref{fig3}. In both figures, $N(\omega)$ [left panels] is shown along with the two spin projections $N_{\uparrow}(\omega)$, $N_{\downarrow}(\omega)$ [right panels]. Spectra in the AF (normal) state are shown as color full lines (grey dashed lines).

For $U=4<U_{\rm MIT}$, Figs.~\ref{fig2}a,b show the DOS for different $T$ across $T_N^d$. At low $T$, $N(\omega)$ [Fig.~\ref{fig2}a] in the AF state has a narrow gap between two prominent peaks, which correspond to the Bogoliubov quasiparticles. The system is an antiferromagnetic {\it insulator} (AF-I). In the infinite-size system, this would be a deep pseudogap~\cite{Vilk:1997}. $U$ is not small enough to be in the static mean-field limit, so one sees small precursors of the Hubbard bands at higher energies~\cite{Preuss1995, Moreo1995, Vilk:1997}. Nevertheless, the behavior is close to that expected in static mean-field: as $T$ increases, the Bogoliubov peaks broaden and the distance between the two peaks decreases, reflecting the closing of the AF gap (Fig.~\ref{fig2}c). Increasing $T$ above $T_N^d$, the Bogoliubov peaks merge as the AF gap closes. The normal-state is metallic and the peak at the Fermi level is what is left from the Van Hove singularity. 
Hence, for $U<U_{\rm MIT}$, the AF-I is born out of a metallic normal state. As expected from mean-field, the spectral weight rearranges itself mostly over a frequency range on the scale of the gap. This is also the case for the spin-projected spectra in Fig.~\ref{fig2}b: even though at low $T$, Bogoliubov peaks are quite spin polarized, the difference between the two spin projections becomes smaller at frequencies above the gap. 

For $U=12>U_{\rm MIT}$ the behavior is qualitatively different. The spin-projected spectra in Fig.~\ref{fig2}e show that the difference between $N(\omega)$ for the two spin projections survives for a huge frequency scale, much larger than the gap size, especially at low $T$. This is typical of strongly correlated systems. That normal and AF state differ over large frequency scales is also apparent in Fig.~\ref{fig2}d. The overall shape of $N(\omega)$ is also different from the weakly correlated case. In the N\'eel state in Fig.~\ref{fig2}d there is a large gap surrounded by a four-peak structure: two Hubbard bands separated by a gap of order $U$, and two Bogoliubov peaks at the lower edges of the Hubbard bands~\cite{Preuss1995, Moreo1995, Vilk:1997}. Hence the system is, as before, an AF {\it insulator}, but it evolves differently with $T$. 
Similarly to $U=4$, the Bogoliubov peaks broaden with increasing $T$, yet the spectral weight shifts from Bogoliubov peaks to Hubbard bands. 
For $T<T_N^d$, $N(\omega)$ has the characteristics of an AF-I. Raising $T$ above $T_N^d$ shows that the AF-I is born out of a Mott insulator. As observed previously~\cite{phk}, even above $T_N^d$, the normal-state $N(\omega)$ displays remnants of the Bogoliubov peaks at the Hubbard bands edges that reflect the short-range AF correlations allowed by the CDMFT solution. 

The contrast between weak and strong interactions is also clear in Figs.~\ref{fig3}a,b that display the evolution of $N(\omega)$ with $U$ at $T<T_{\rm MIT}<T_N^d$. The Mott transition is visible between $U=5$ and $U=6$ in the normal-state $N(\omega)$. 
At low $U$, only the two Bogoliubov peaks are present in $N(\omega)$ while at large $U$, only the two featureless Hubbard bands are present. At intermediate $U$, $N(\omega)$ has a characteristic four-peak structure: two Bogoliubov peaks, which flank two Hubbard bands or their precursors~\cite{Preuss1995, Moreo1995, Vilk:1997}. These structures evolve in a qualitatively different way with $U$: (i) the Bogoliubov peaks are quite narrow below $U_{\rm MIT}$, progressively broaden with increasing $U$ and dissolve into featureless Hubbard bands at $U\gg U_{\rm MIT}$. Their maximum value occurs just below $U_{\rm MIT}$; (ii) the precursors of the Hubbard bands appear for $U\lessapprox U_{\rm MIT}$ and evolve into well defined Hubbard bands for $U\gtrapprox  U_{\rm MIT}$. A comparison between the AF and normal $N(\omega)$ reveals that, for $U>U_{\rm MIT}$, the AF gap $\Delta_{\rm AF}$ is larger than the Mott gap or its precursors $\Delta_M$, similarly to the 3D case in Ref.~\onlinecite{WangAF:2009} (see Fig.~\ref{fig3}c,d).

\section{Conclusions}
\label{concl}
The crossover between potential- and kinetic-energy driven antiferromagnetism contains clues on the mechanism of antiferromagnetism and, contrary to the superconducting case~\cite{Chester:1956, LeggettCondEn, scalapinoCondEn, Norman:2000, Molegraaf2002, deutscher2005, andersonBOOK, Hirsch1, Demler1998, maierENERGY, carbone2006, kyungPAIRING, millisENERGY, ssht, LorenzoSC}, has been largely overlooked up to now. For the 2D Hubbard model, we addressed this problem and revealed distinctive features of the double occupancy, potential, kinetic and total energies, and local DOS in two phases, normal and antiferromagnetic. The underlying Mott transition and its associated Widom line leave their mark on the AF phase through sharp crossovers  associated with them. 
Thus we demonstrated that it is possible for the Mott transition to determine complex changes in observables associated with the AF phase. 

Although a crossover between weakly and strongly correlated antiferromagnetism is expected, our work goes beyond simple expectations: our findings add depth and understanding, and could not have been anticipated on general grounds. Our detailed mapping of the $U$-temperature phase diagram allowed us to identify observable changes in behavior occuring along a line that extends from zero temperature along the Mott transition and the associated Widom line, demonstrating the importance of the underlying normal state in determining properties of the ordered state. 
Specifically we found the following: 
(i) A rich behavior of the difference in kinetic, potential, and total energy between the AF and normal state: at low enough temperature, we found that the critical $U$ at which $\Delta E_{\rm pot}$ and $\Delta E_{\rm kin}$ cross zero, and $\Delta E_{\rm tot}$ is largest, is determined by the first-order Mott transition (cf Fig.~\ref{fig1}d). For $T>T_{\rm MIT}$, the change in energetics is even more surprising: The region where $\Delta E_{\rm pot}(T)$ crosses zero follows the loci of the largest condensation energy, which in turn correlates with the normal state Widom line emanating out of the Mott endpoint (cf Fig.~\ref{fig1}e). Note that there is no reason to expect they should be right on, as demonstrated by the behavior of the difference in kinetic energy between normal and AF state, which departs from it (see Fig.~\ref{figSM6} in Appendix~\ref{Sec:Energy}).  
(ii) A distinctive behavior of the double occupancy $D$ versus temperature. As shown in Figs.~\ref{fig1}b,c, the AF state for $U < U_{\rm MIT}$ suppresses $D$ compared with $D$ in the normal state, and reverses the slope of $D(T)$. On the other hand, for $U > U_{\rm MIT}$, $D$ is increased in the AF state. Our benchmarks of double occupancy with alternative methods (cf Fig.~\ref{fig1}b,c) also provide insight into the meaning of such calculations. In particular, we found that double occupancy in the low temperature ordered state is closer to the correct result than if it had been computed in the normal state. In fact, the temperature dependence is qualitatively correct only if the CDMFT solution is in the AF phase. 
(iii) Striking differences in the local DOS at small/large $U$. As shown in Figs.~\ref{fig2}, \ref{fig3}, the frequency range over which AF rearranges spectral weight and the frequency-dependence of spin polarized spectra are surprising. In particular, the fact that the spectrum is rearranged over energy scales much larger than the AF gap when $U > U_{\rm MIT}$, by contrast with the $U < U_{\rm MIT}$ case, is an important result, characteristic of broken-symmetry gaps opening on strongly-correlated states. Also, the relation between the size of the AF gap and the Mott gap as a function of $U$ is seen in a new light when the full $U$ dependence is plotted, as in Fig.\ref{fig3}a. 

From a broader perspective, our findings open the road to understanding sharp crossovers within other phases or models by careful consideration of normal-state properties. This may be especially fruitful away from half filling where high-temperature superconductivity occurs. Note that at large $U$ and half filling, the kinetic energy gain makes the AF stable (see Fig.~\ref{fig1}d,e). Similarly, at large $U$ and finite hole-doping, it is again the kinetic energy gain that makes the superconductivity stable~\cite{LorenzoSC}. Hence, in both cases, the normal state has an excess of kinetic energy: antiferromagnetism and superconductivity are two phases that reduce it. 

Our findings on the behavior of the double occupancy in the $U$-temperature space, and hence the specific link between Slater/Heisenberg crossover and the underlying Mott transition, could also be tested experimentally in ultra-cold atoms in optical lattices, due to recent experimental advances in quantum gas microscopy~\cite{Greif:2013, Hart:2015, Parsons:2016, Boll:2016, Cheuk:2016}.

\begin{acknowledgments}
This work has been supported by the Natural Sciences and Engineering Research Council of Canada (NSERC) under grant RGPIN-2014-04584, and by the Research Chair in the Theory of Quantum Materials. Simulations were performed on computers provided by the Canadian Foundation for Innovation, the Minist\`ere de l'\'Education des Loisirs et du Sport (Qu\'ebec), Calcul Qu\'ebec, and Compute Canada.
\end{acknowledgments}

\appendix

\section{Method}
\label{Sec:Method}

In this appendix we show some details of the method. The effective action of the impurity (cluster in a bath) problem is given by: 
\begin{align}
S_{\rm eff} = S_{\rm cl}(c, c^\dagger) + \int_0^\beta d\tau d\tau' \, c^\dagger(\tau) \Delta(\tau - \tau') c(\tau'). 
\label{eq:Seff}
\end{align}
To solve the impurity problem, i.e. to find $G_{\rm cl} = G_{\rm cl}[\Delta] = \langle c c^\dagger \rangle_{S_{\rm eff}}$, we use the continuous-time quantum Monte Carlo method based on the expansion of $S_{\rm eff}$ in the hybridisation function (CT-HYB) $\Delta$~\cite{millisRMP}. 
In this section we shall show that, especially in the antiferromagnetic (AF) phase, a judicious analysis of the symmetries of the problem have important consequences on choice of the single-particle basis in Eq.~\ref{eq:Seff}, on the ergodicity of the CT-HYB impurity solver, and on the Monte Carlo sign problem. For details on convergence and Monte Carlo error bars, see Ref.~\onlinecite{patrickCritical}

\subsection{Symmetries}
Two remarks about symmetries are in order: (a) The cellular dynamical mean-field theory (CDMFT) self-consistent mapping onto the impurity problem preserves the symmetries of the lattice system compatible with the partitioning (here, the symmetries of a $2\times2$ plaquette). In the normal phase, the dynamical mean field $\Delta$ is constrained to satisfy these symmetries, whereas in the N\'eel state, some of the above symmetries are broken. (b) Using the symmetries of the impurity Hamiltonian allows one to speed up the calculations~\cite{hauleCTQMC}. We thus choose a single-particle basis for $c$ (or $c^+$) that transform according to the irreducible representations of an Abelian point group that represent the spatial symmetries of the impurity Hamiltonian. The Abelian group chosen must be a subgroup of the total point group of the impurity Hamiltonian.

These remarks suggest the following implementation.

\begin{figure}
\centering{
\includegraphics[width=0.4\linewidth,clip=]{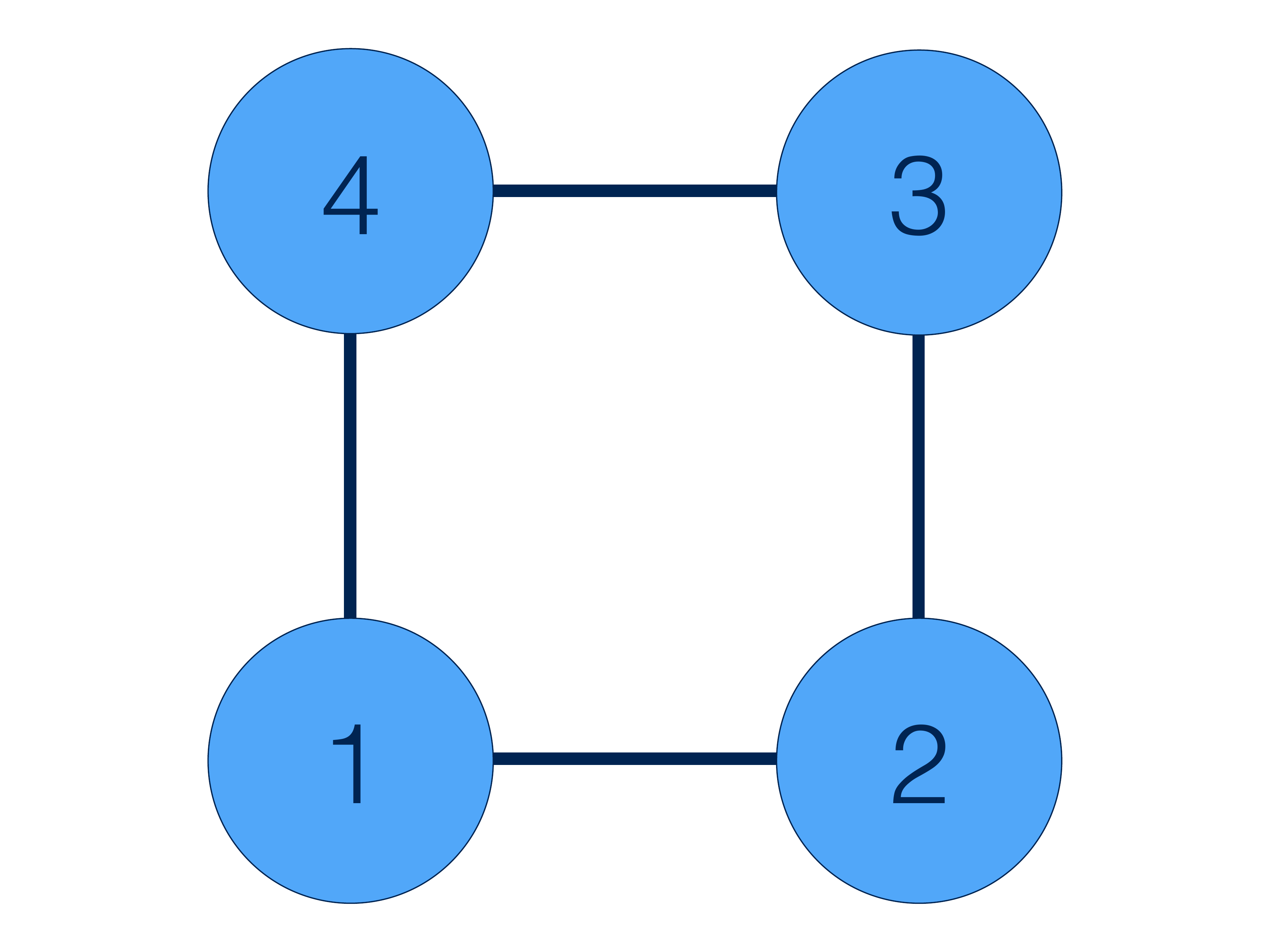}}
\caption{$2\times 2$ plaquette for the CDMFT. }
\label{figS1}
\end{figure}

\subsubsection{Normal state}
The {\it normal state} satisfies (i) charge conservation ($U(1)$) symmetry, (ii) time-reversal symmetry, (iii) spin rotational ($SU(2)$) symmetry, (iv) translational symmetry, and (v) point group $C_{4v}$ symmetry of the plaquette. CDMFT breaks translational symmetry but it is possible to satisfy all other symmetries. To speed up calculations of the trace for operators on the cluster,~\cite{hauleCTQMC} it is convenient to classify cluster states using a point group with Abelian symmetry. We choose the point group $C_{2v}$ with mirrors along horizontal and vertical axis of the plaquette. The appropriate basis in the irreducible representations $A_1$, $A_2$, $B_1$, $B_2$ of $C_{2v}$, is (see Fig.~\ref{figS1} for indices): 
\begin{align}
c_{A_1\sigma} &= \tfrac{1}{2}(c_{1\sigma} + c_{2\sigma} + c_{3\sigma} + c_{4\sigma}) \label{a1} \\
c_{A_2\sigma} &= \tfrac{1}{2}(c_{1\sigma} - c_{2\sigma} + c_{3\sigma} - c_{4\sigma})  \label{a2} \\
c_{B_1\sigma} &= \tfrac{1}{2}(c_{1\sigma} + c_{2\sigma} - c_{3\sigma} - c_{4\sigma}) \label{b1} \\
c_{B_2\sigma} &= \tfrac{1}{2}(c_{1\sigma} - c_{2\sigma} - c_{3\sigma} + c_{4\sigma}). \label{b2}
\end{align}
Such a choice gives, for each spin $\sigma$, a $4\times 4$ diagonal hybridisation function matrix $\Delta$,
\begin{align}
\Delta_{\sigma,\sigma} & = \left( \begin{array}{cccc} 
\Delta_{A_1 \sigma , A_1 \sigma } & 0 & 0 & 0  \\ 
0 & \Delta_{A_2 \sigma , A_2 \sigma} & 0 & 0  \\ 
0 & 0 &  \Delta_{B_1 \sigma , B_1 \sigma } & 0  \\ 
0 & 0 & 0 &  \Delta_{B_2 \sigma , B_2 \sigma } 
\end{array}\right). 
\end{align}
To enforce time-reversal symmetry (i.e. to satisfy requirement (ii)), we constrain up and down spins to take same values. While we limited ourselves to the Abelian $C_{2v}$ group in the choice of the irreducible representations, we can still make use of the $C_{4}$ rotation symmetry, by imposing that
\begin{equation}
\Delta_{B_1 \sigma,  B_1 \sigma} = \Delta_{B_2 \sigma , B_2 \sigma }. 
\end{equation}
There are thus only three independent hybridisation functions in the normal state.

\subsubsection{N\'eel antiferromagnetic state}
By contrast, on the $2\times 2$ plaquette, the {N\'eel antiferromagnetic phase} breaks (i) time-reversal symmetry, (ii) spin rotational symmetry, and (iii) $C_4$ ($\pi/2$) rotation. However, time reversal combined with $C_4$ is still a symmetry of this system. Even though spin-rotational symmetry is broken, the total component of spin along the $z$ direction is preserved in both the normal and antiferromagnetic states. Since we are not interested in expectation values in directions other than the $z$ direction, the breaking of spin-rotational symmetry is inconsequential. 

While AF does break spatial $C_{2v}$ symmetry with mirrors along horizontal and vertical axis of the plaquette, $C_{2v}$ point-group symmetry {\it with mirrors along the diagonals} is preserved. This suggests to work in the single-particle basis (see Fig.~\ref{figS1} for indices):
\begin{align}
c_{A_1\sigma} &= \tfrac{1}{\sqrt{2}}
(c_{1\sigma} + c_{3\sigma}) \label{Da1} \\
c_{A'_1\sigma} &= \tfrac{1}{\sqrt{2}}
(c_{2\sigma} + c_{4\sigma}) \label{Da1p} \\
c_{B_1\sigma} &= \tfrac{1}{\sqrt{2}}
(c_{1\sigma} - c_{3\sigma}) \label{Db1} \\
c_{B_2\sigma} &= \tfrac{1}{\sqrt{2}}
(c_{2\sigma} - c_{4\sigma}). \label{Db2}
\end{align}
This basis gives, for each spin, a $4\times 4$ block-diagonal hybridisation function matrix $\Delta$, with one $2\times2$ block ($A_1$) and two $1\times1$ blocks ($B_1$ and $B_2$). For spin $\sigma$ it takes the form: 
\begin{align}
\Delta_{\sigma,\sigma} & = \left( \begin{array}{cccc} 
\Delta_{A_1 \sigma , A_1 \sigma } & \Delta_{A_1 \sigma , A'_1 \sigma } & 0 & 0  \\ 
\Delta_{A'_1 \sigma , A_1 \sigma } & \Delta_{A'_1 \sigma , A'_1 \sigma} & 0 & 0  \\ 
0 & 0 &  \Delta_{B_1 \sigma , B_1 \sigma } & 0  \\ 
0 & 0 & 0 &  \Delta_{B_2 \sigma , B_2 \sigma } 
\end{array}\right). 
\end{align}
There are five nonzero independent imaginary-time hybridisation functions in the above matrix. Indeed, since the imaginary-time hybridisation function is real, we have 
\begin{equation}
\Delta_{A_1 \sigma,  A'_1 \sigma} = \Delta_{A'_1 \sigma , A_1 \sigma }. 
\end{equation}

There is no additional independent hybridisation function for spins $\bar{\sigma}=-\sigma$. Indeed, symmetry under rotation by $\pi/2$ followed by a spin flip transforms the operators as follows
\begin{align}
& c_{A_1\sigma} \rightarrow  c_{A'_1\bar{\sigma}} \\
& c_{A'_1\sigma} \rightarrow c_{A_1\bar{\sigma}} \\
& c_{B_1\sigma} \rightarrow c_{B_2\bar{\sigma}} \\
& c_{B_2\sigma} \rightarrow - c_{B_1\bar{\sigma}}, 
\end{align}
which in turn implies for the down-spin hybridisation function
\begin{align}
\Delta_{\bar\sigma,\bar\sigma} & = \left( \begin{array}{cccc} 
\Delta_{A'_1 \sigma , A'_1 \sigma } & \Delta_{A'_1 \sigma , A_1 \sigma } & 0 & 0  \\ 
\Delta_{A_1 \sigma , A'_1 \sigma } & \Delta_{A_1 \sigma , A_1 \sigma} & 0 & 0  \\ 
0 & 0 &  \Delta_{B_2 \sigma , B_2 \sigma } & 0  \\ 
0 & 0 & 0 &  \Delta_{B_1 \sigma , B_1 \sigma } 
\end{array}\right). 
\end{align}
In the normal state there are only three independent hybridisation functions, as in the previous section, since symmetry implies the additional equalities for both spin species
\begin{align}
\Delta_{A'_1 \sigma , A'_1 \sigma }&=\Delta_{A_1 \sigma , A_1 \sigma } \label{h_diag_a} \\ 
\Delta_{B_1 \sigma , B_1 \sigma }&=\Delta_{B_2 \sigma , B_2 \sigma }  \label{h_diag_b}. 
\end{align} 

\subsection{Ergodicity}
Reference~\onlinecite{patrickERG} demonstrates that for several classes of broken symmetries that involve spatial components, CT-HYB impurity solver is not ergodic as a matter of principle if one follows the usual updating procedure that adds or removes a  pair of creation-annihilation operators (``2-operator updates''). Quite generally, if the cluster has more symmetries than the bath, it may be necessary to make updates with larger numbers of pairs of creation-annihilation operators (``n-operator updates'') to cure the lack of ergodicity. As an example~\cite{patrickERG}, 4-operator updates are necessary for ergodicity when one considers $d$-wave superconductivity within $2\times 2$ CDMFT. The problem in that case is as follows. The expansion in powers of the hybridisation contains products of two symmetry-breaking hybridisation functions whose associated cluster operators, four of them, recover the full symmetry and hence lead to non-vanishing traces in the cluster, while a single pair of operators associated with one of these hybridization functions leads to a vanishing trace because of the broken symmetry. Hence, the four-operator non-vanishing contributions to the trace cannot appear as a sequence of updates involving only pairs of operators. 

Here, in the antiferromagnetic state, the choice of the  $C_{2v}$ group symmetry with mirrors along horizontal and vertical axes does not leave the Hamiltonian of the bath invariant in the AF state, but it leaves the Hamiltonian of the cluster invariant. Hence, more than 2-operator  updates would be necessary to attain ergodicity if that choice was made. On the other hand, the cluster and the bath have the same $C_{2v}$ point-group symmetry {\it with mirrors along the diagonals}. That is the point-group symmetry whose irreducible representations we use as a basis. The usual updates that add or remove a pair of creation-annihilation operators then do not couple to point-group symmetries that exist only in the cluster and not in the bath and they therefore suffice for ergodicity. 
Therefore, with such a choice of point-group symmetry, the usual updates that add or remove a pair of creation-annihilation operators are ergodic. Note that the updates always preserve total $S_z=0$, which is a symmetry of both phases. 

\begin{figure}
\centering{
\includegraphics[width=0.9\linewidth,clip=]{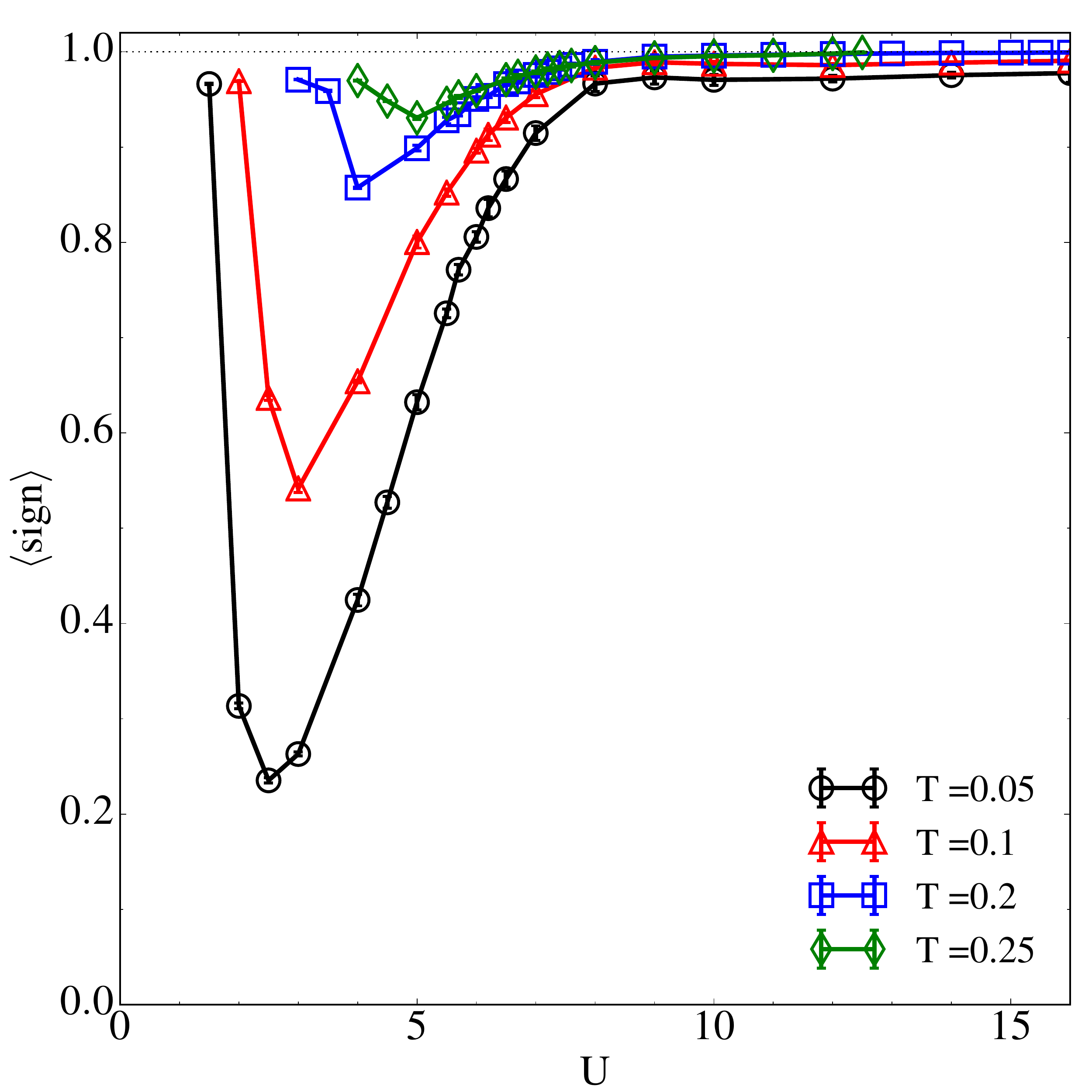}
\includegraphics[width=0.9\linewidth,clip=]{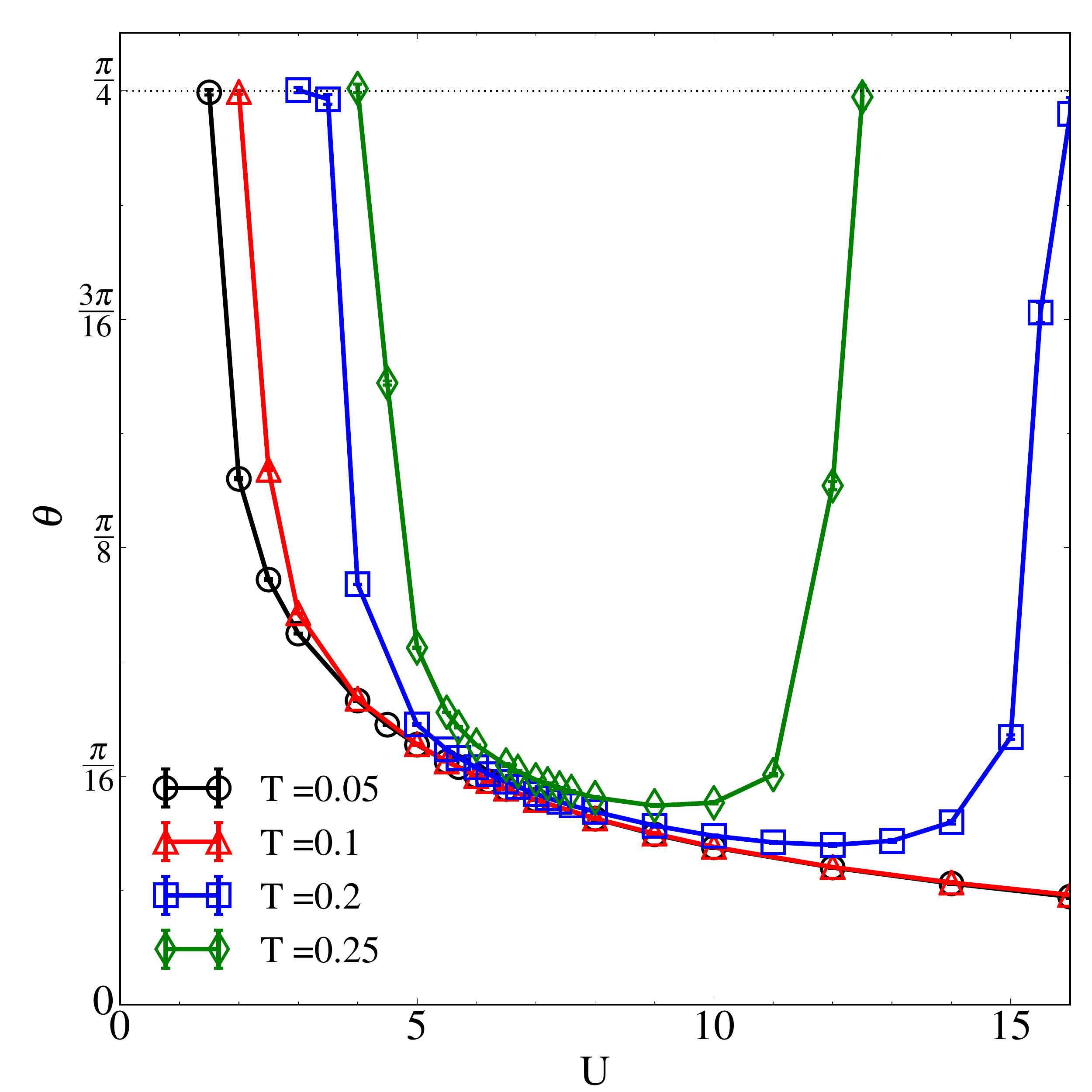}
}
\caption{(a) Average sign in CTQMC AF simulations versus $U$. (b) Angle $\theta$ in Eqs.~\ref{eq:theta1}, \ref{eq:theta2} versus $U$. Data are shown for temperatures $T=1/20, 1/10, 1/5$ and $1/4$. The angle $\pi/4$ appears only in the normal phase, namely when $m_z=0$.}
\label{figSergodicity}
\end{figure}

\subsection{Sign problem and addendum on ergodicity}
Generally, if we compare the sign problem for identical values of $U$ and $T$ in the normal and antiferromagnetic phase, the sign problem in the antiferromagnetic phase is worse. As pointed out in Ref.~\onlinecite{patrickCritical}, to mitigate the sign problem, one can exploit the degeneracy in the $A_1$ subspace by introducing an angle $\theta$ in the choice of basis:  
\begin{align}
c_{A_1\sigma} &= \tfrac{\cos\theta}{\sqrt{2}}
(c_{1\sigma} + c_{3\sigma}) +
\tfrac{\sin\theta}{\sqrt{2}}(c_{2\sigma} + c_{4\sigma}) \label{eq:theta1} \\
c_{A'_1\sigma} &= \tfrac{\sin\theta}{\sqrt{2}}
(c_{1\sigma} + c_{3\sigma}) -
\tfrac{\cos\theta}{\sqrt{2}}(c_{2\sigma} + c_{4\sigma}) .
\label{eq:theta2}
\end{align}
Figure~\ref{figSergodicity} shows the average sign (top panel) and angle $\theta$ (bottom panel) as a function of $U$ for different temperatures. The usual basis, Eqs.~\ref{Da1}-\ref{Db2}, is recovered by setting $\theta=0$, but it gives a bad sign problem. One can choose the appropriate $\theta$ in order to minimize the sign problem~\cite{patrickCritical}. $\theta$ must be corrected at each iteration of the CDMFT self-consistency loop in order to minimize the off-diagonal component of the $2\times2$ block ($A_1$) of the hybridization function matrix. 

This procedure of changing $\theta$ to minimize the sign problem was used in the AF state only. For the normal state, it is $\theta=\pi/4$ that minimizes the sign problem. Note that for the AF phase, the choice $\theta=\pi/4$ corresponds to a change from the $A_1$ and $A'_1$ basis to the basis that transforms like the $A_1$ and $A_2$ representations, Eqs.~\eqref{a1} and \eqref{a2}, of the $C_{2v}$ symmetry with mirrors along horizontal and vertical axis. For that particular angle then, one encounters the ergodicity problem~\cite{patrickERG} mentioned above. Indeed, the hybridization functions $ \Delta_{A_1 \sigma,  A_2 \sigma}$ and $ \Delta_{A_2 \sigma , A_1 \sigma }$ do not vanish in the antiferromagnetic state. When the product of these two functions appears in the hybridization expansion, the corresponding product of operators in the cluster, four of them, does not vanish. However, the products of operators corresponding to an odd number of $ \Delta_{A_1 \sigma,  A_2 \sigma}$ or $ \Delta_{A_2 \sigma , A_1 \sigma }$ vanishes because the cluster does not break the symmetry while the product of $A_1$ and $A_2$ does. Therefore ergodicity is not attained with 2-operator updates only, as explained in Sec.~\ref{Sec:Method}.B. As shown in Fig.~\ref{figSergodicity}b, we verified that, as long as the AF phase is sustained by CDMFT equations, i.e. as long as the staggered magnetisation $m_z$ is nonzero (see Fig.~1 of main text and Fig.~\ref{figS2} in Appendix~\ref{Sec:mz}), the angle $\theta$  differs from $\pi/4$ and there is no ergodicity problem.

\section{Order parameter}
\label{Sec:mz}

\begin{figure}[t!]
\centering{
\includegraphics[width=0.95\linewidth,clip=]{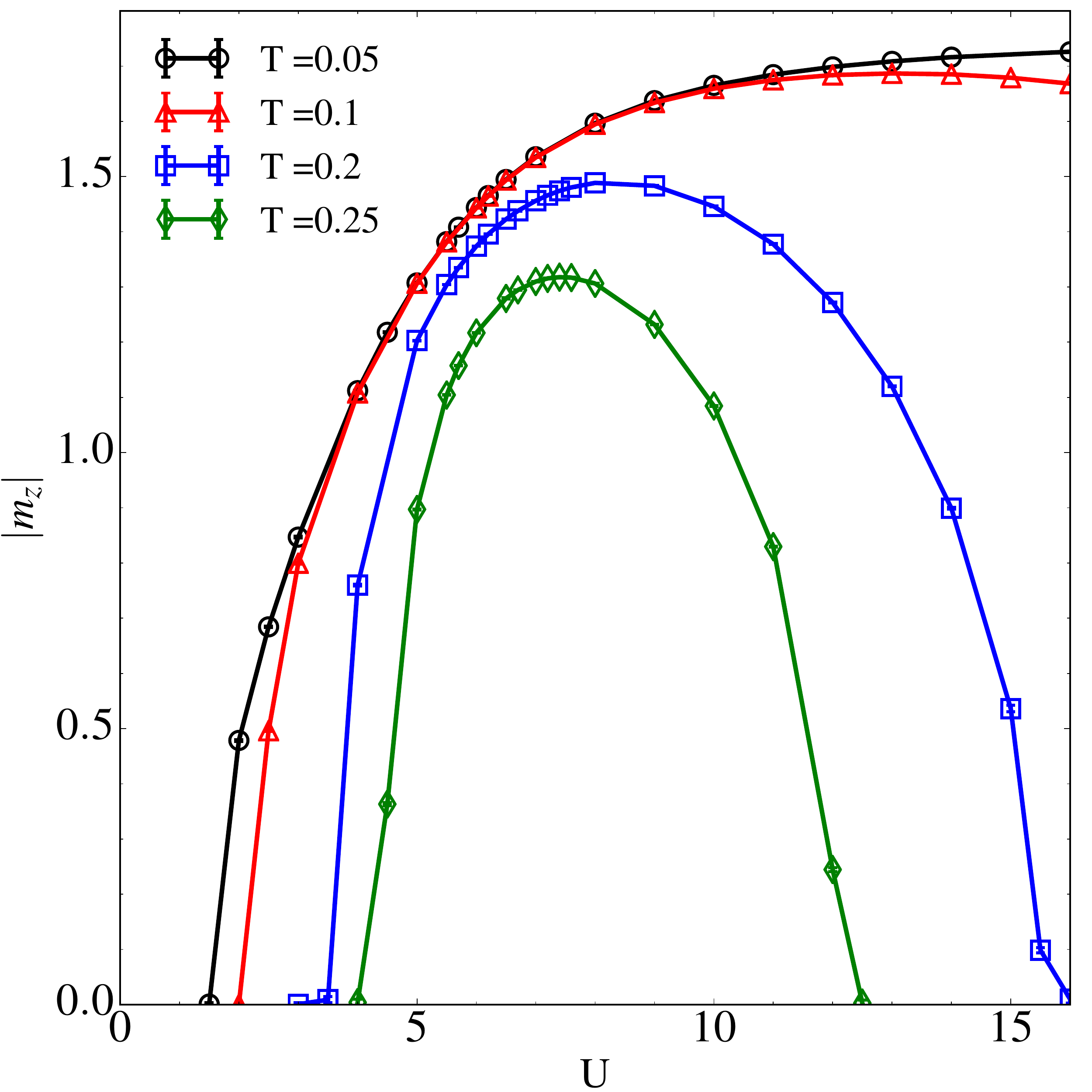}}
\caption{Staggered magnetisation $m_z$ versus $U$ for different values of $T$.  $T^d_N$ is obtained from the mean of the two temperatures where $m_z$ changes from finite to a small value (here $m_z =0.045$). }
\label{figS2}
\end{figure}

In this appendix we present the raw data of the staggered magnetisation $m_z$, to complement the colormap in Figure~\ref{fig1}a of the main text. To determine the parameter space where AF arises from the CDMFT equations, we determine where the staggered magnetisation
\begin{align}
m_z & = \frac{2}{N_c} \sum_i(-1)^i (n_{i\uparrow} - n_{i\downarrow})
\end{align}
is nonzero.  
Figure~\ref{figS2} shows $m_z$ as a function of $U$ for different temperatures. These scans lead to the colormap in Figure~1a of the main text.  

\section{Energetics}
\label{Sec:Energy}

\begin{figure*}[ht!]
\centering{
\includegraphics[width=0.999\linewidth,clip=]{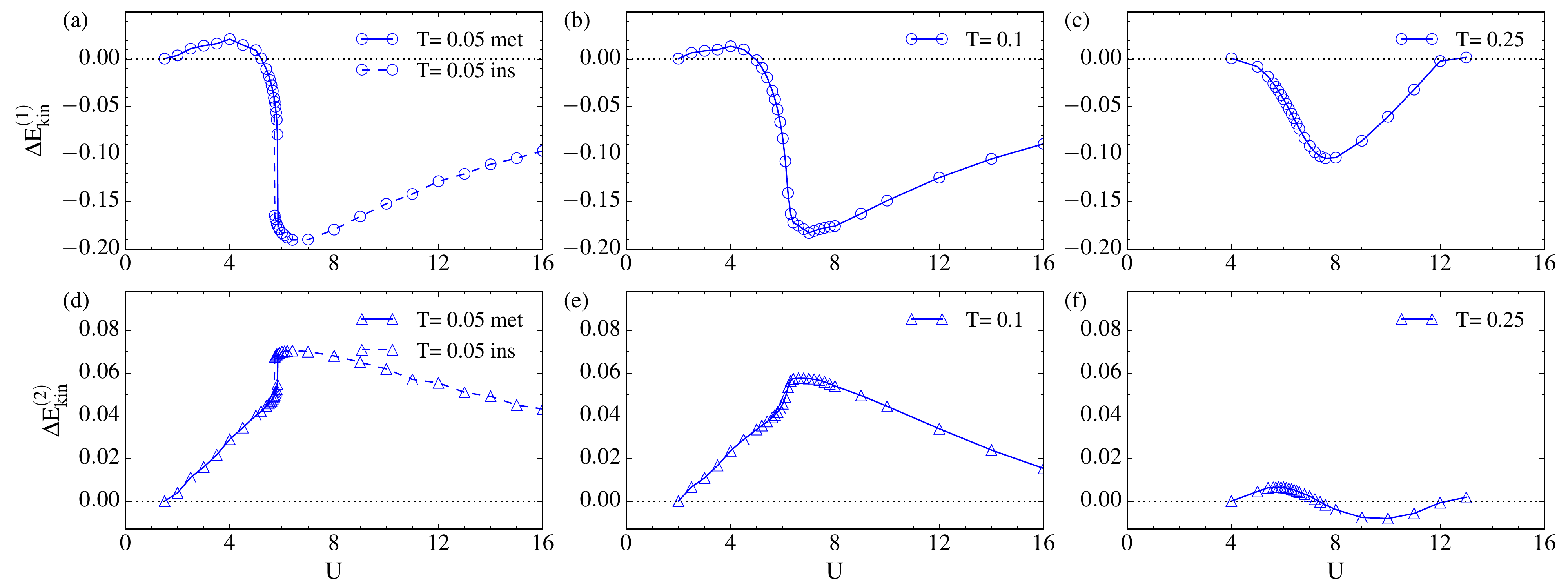}}
\caption{Contributions to the difference in kinetic energy between AF and normal state, as a function of $U$. Top (bottom) panels show $\Delta E_{\rm kin}^{(1)}$ ($\Delta E_{\rm kin}^{(2)}$). Data are shown for $T=1/20<T_{\rm MIT}$, $T=1/10>T_{\rm MIT}$,  and $T=1/4$ (left, central and right columns, respectively). For $T<T_{\rm MIT}$ two normal state solutions coexist at the Mott transition.}
\label{figSM3}
\end{figure*}
\begin{figure*}
\centering{
\includegraphics[width=0.999\linewidth,clip=]{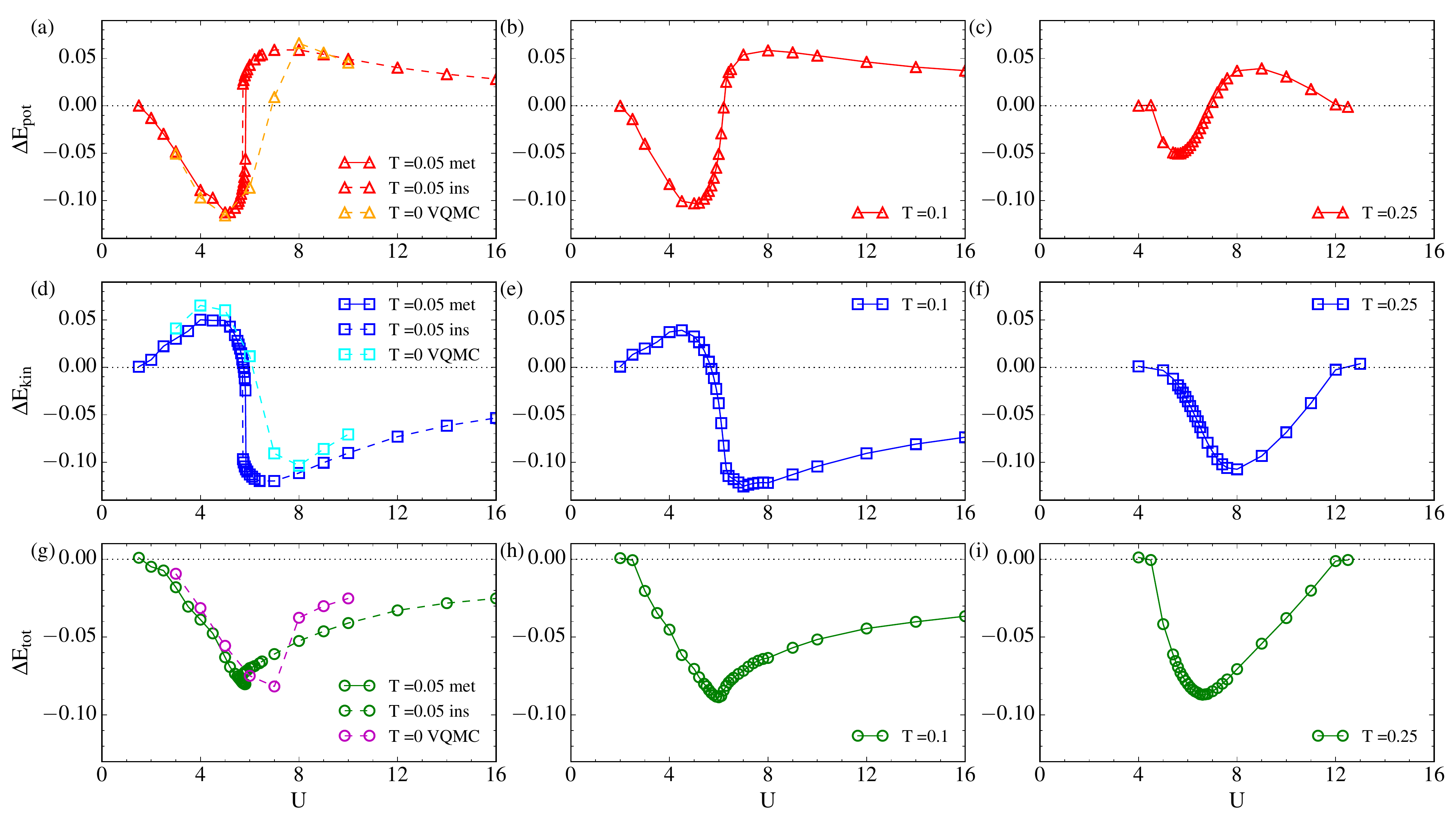}}
\caption{Difference in potential energy $\Delta E_{\rm pot} = E_{\rm pot}^{\rm AF} -E_{\rm pot}^{\rm NS}$ (top panels), in kinetic energy $\Delta E_{\rm kin}$ (central panels) and in total energy $\Delta E_{\rm tot}$ (bottom panels) as a function of $U$. As in Fig.~\ref{figSM3}, data are shown for $T=1/20<T_{\rm MIT}$, $T=1/10>T_{\rm MIT}$,  and $T=1/4$ (left, central and right columns, respectively). For benchmark, at our lowest temperature (panels a,d,g) we show the $T=0$ variational QMC calculation of Ref.~\onlinecite{Tocchio2016}. 
}
\label{figSM4}
\end{figure*}

In this appendix we elaborate on the energetics of the model. 
For CDMFT solved with CT-HYB, we demonstrated in Ref.~\onlinecite{LorenzoSC} that the kinetic energy is the sum of two contributions, $E_{kin} =E_{kin}^{(1)} +E_{kin}^{(2)}$, where $E_{kin}^{(1)}$ is a term related to the average expansion order $\langle k \rangle$: 
\begin{align}
E_{kin}^{(1)} & = -\frac{\langle k \rangle}{N_c \beta}
\end{align}
and $E_{kin}^{(2)}$ is a term related to the cluster part: 
\begin{align}
E_{kin}^{(2)} & = \frac{2T}{N_c} \sum_n e^{-i\omega_n0^-} \sum_{ij}\left[ t_{ij}^{imp} G_{ji}^{imp}(i\omega_n) \right], 
\end{align}
where $N_c$ is the cluster size (here $N_c=4$), $\beta$ is the inverse temperature, $t_{ij}^{imp}$ and $G_{ij}^{imp}$ are the hopping and the Green's function of the impurity problem, respectively. 

Figure~\ref{figSM3} shows these two contributions to the difference in kinetic energy between AF and normal state, as a function of $U$ and $T=1/20, 1/10, 1/4$ (left, central and right panels). 
Figure~\ref{figSM4} displays the difference in kinetic, potential, total energy between AF and normal state versus $U$ and $T=1/20, 1/10, 1/4$ (left, central and right panels). The data at the lowest temperature ($T=1/20$) are shown in Fig.~1d of the main text. For benchmark, in Figs.~\ref{figSM4}a,d,g we also show the $T=0$ variational QMC calculation of Ref.~\onlinecite{Tocchio2016}. 
 
\begin{figure}
\centering{
\includegraphics[width=0.995\linewidth,clip=]{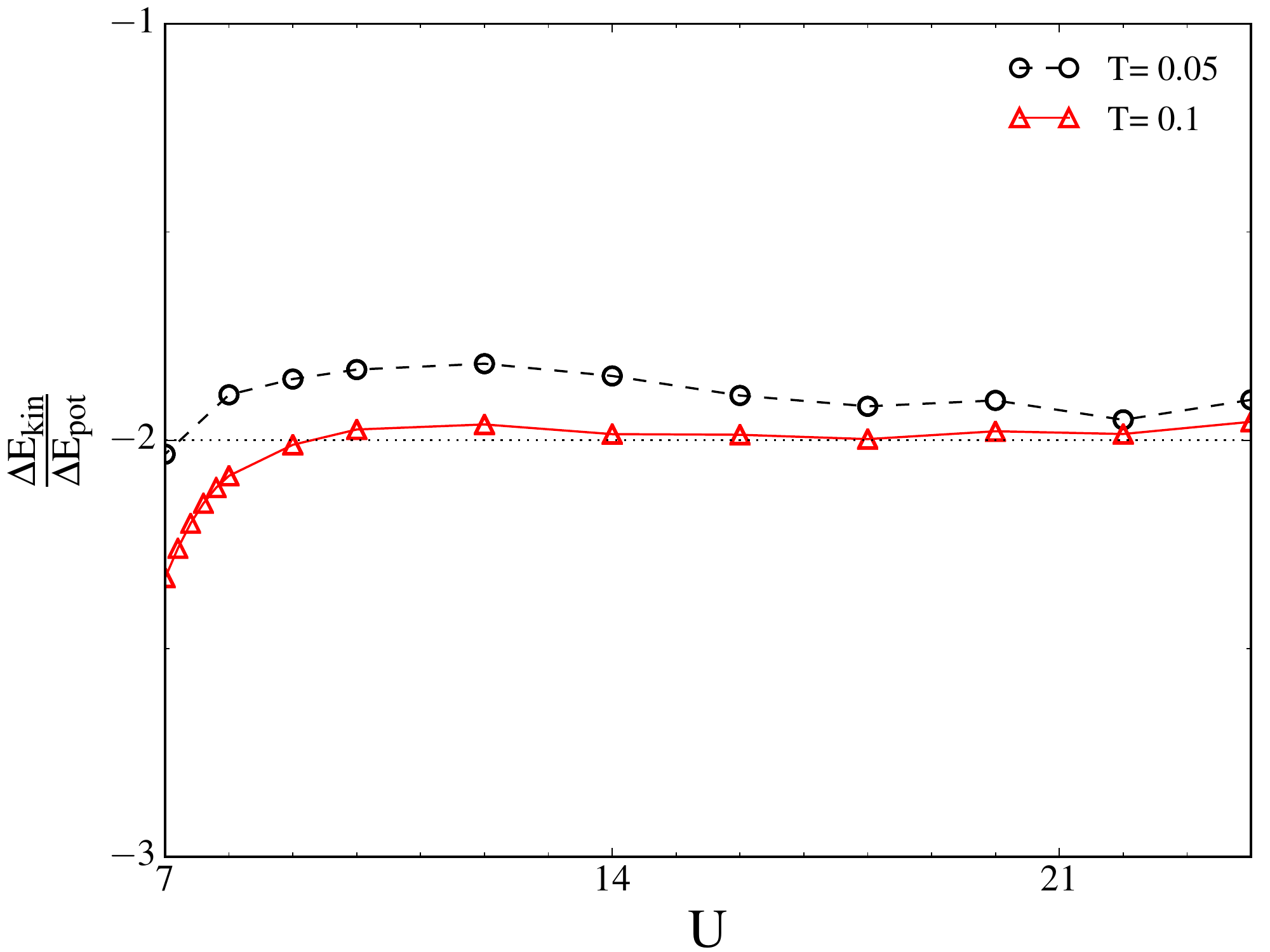}}
\caption{Ratio between kinetic energy gain and potential energy gain versus $U$, at $T=1/10$ (red triangles) and $T=1/20$ (black circles). At large $U$, $\Delta E_{\rm kin}$ is approaching minus twice $\Delta E_{\rm pot}$.}
\label{figSM5}
\end{figure}

Figure~\ref{figSM5} shows that the ratio between the kinetic energy gain and potential energy gain is approaching $-2$ for large $U$, as expected~\cite{FazekasBook:1999} from the exchange energy proportional to $J$. 

\begin{figure}
\centering{
 \includegraphics[width=0.995\linewidth,clip=]{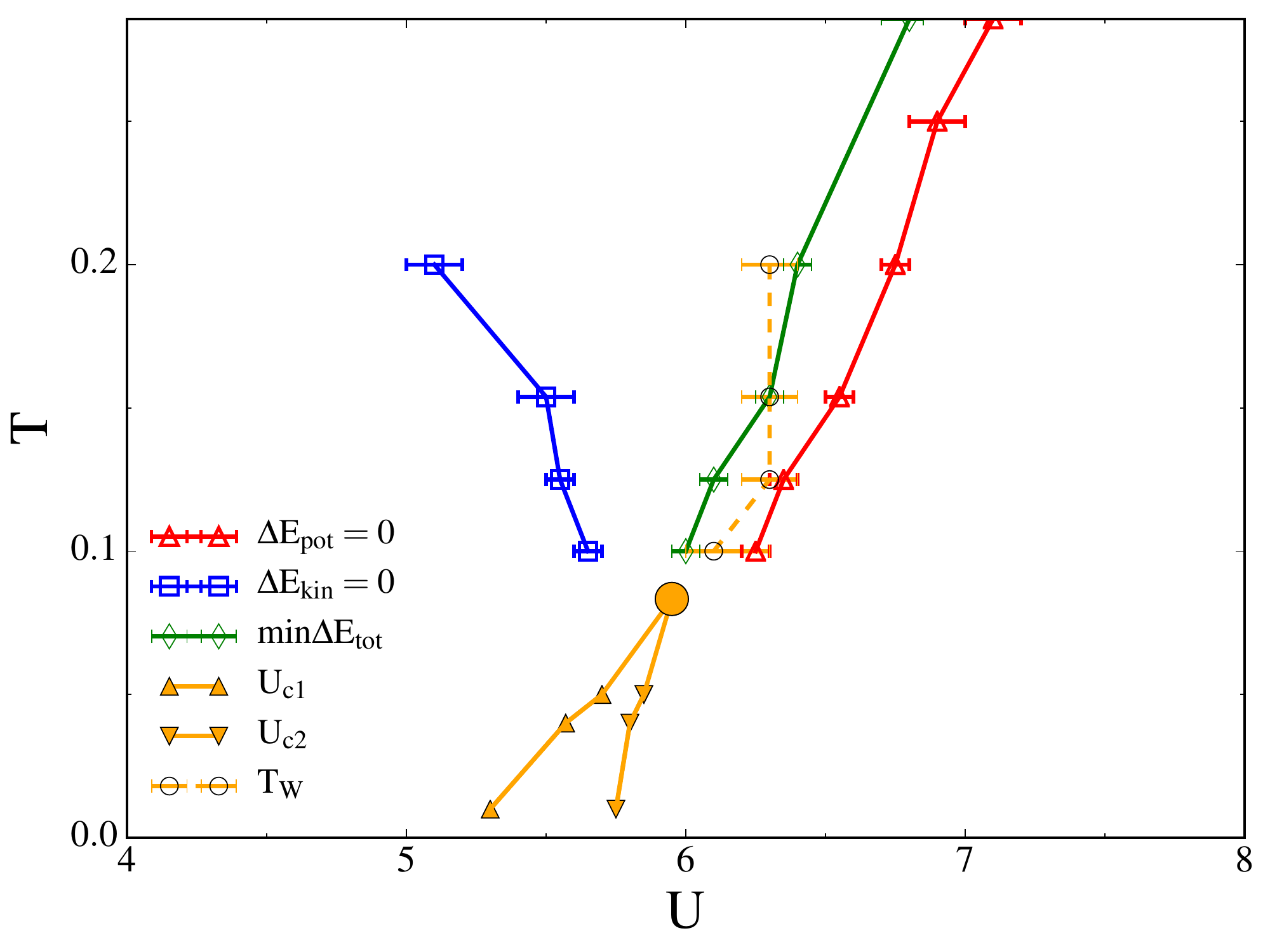}}
\caption{Phase diagram $T-U$ of the two-dimensional half-filled Hubbard model solved with plaquette CDMFT. This figure extends Fig.~\ref{fig1}e of the main text. In the normal state, the first-order Mott metal-insulator transition is delimited by the spinodal lines $U_{c1}(T)$ and $U_{c2}(T)$ (up and down triangles, respectively), and terminates at the critical endpoint (filled orange circle). The Widom line $T_{\rm W}$ (open circles with dotted orange line) is a crossover that extends the first-order transition in the supercritical region and is determined by the inflection points along paths at constant temperature in the double occupancy $D$ (i.e. by max $dD/dU|_T$). 
At values of $U$ and $T$ where the properties of the underlying normal state are governed by the Mott transition we find sharp crossovers between weakly and strongly correlated AF. The loci where the potential (kinetic) energy changes sign are shown by red triangles (blue squares). The line where $\Delta E_{\rm pot}$ changes sign extends up to the maximum of the AF dome. It parallels the loci of the largest condensation energy (green diamonds), which in turn correlates with the Widom line $T_{\rm W}$. The line where $\Delta E_{\rm kin}$ changes sign  still emerges from the Mott endpoint, but extends down to $U\approx 5$ and up to $T\approx 1/5$. 
}
\label{figSM6}
\end{figure}

Finally, Figure~\ref{figSM6} sums up our results in the $T-U$ phase diagram.

\section{Density of states}
\label{Sec:DOS}

In this appendix we show the density of states for different values of $U$, to extend data displayed in Figures~\ref{fig3}a,b of the main text.

\begin{figure*}[ht!]
\centering{
\includegraphics[width=0.45\linewidth,clip=]{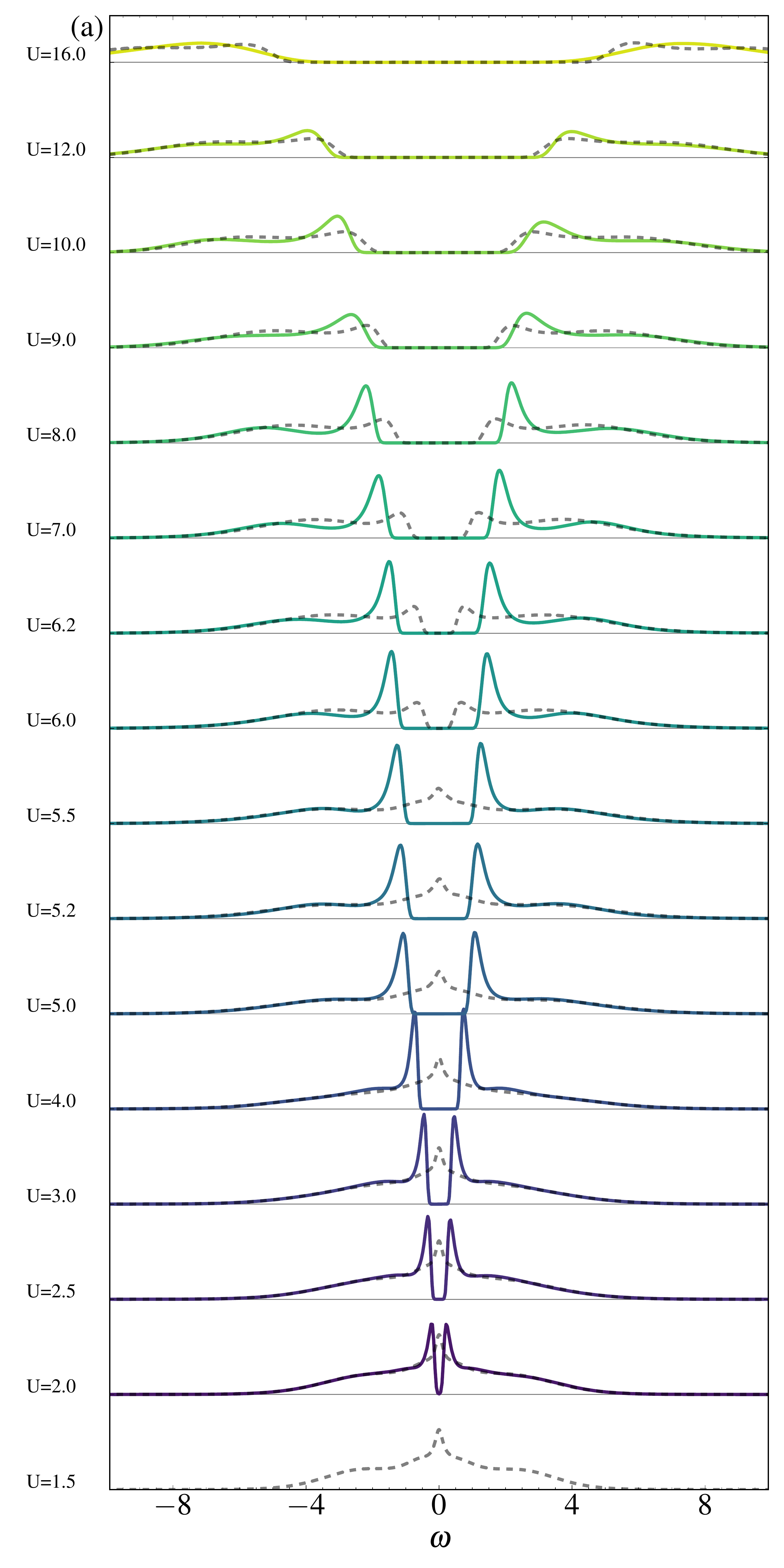}
\includegraphics[width=0.45\linewidth,clip=]{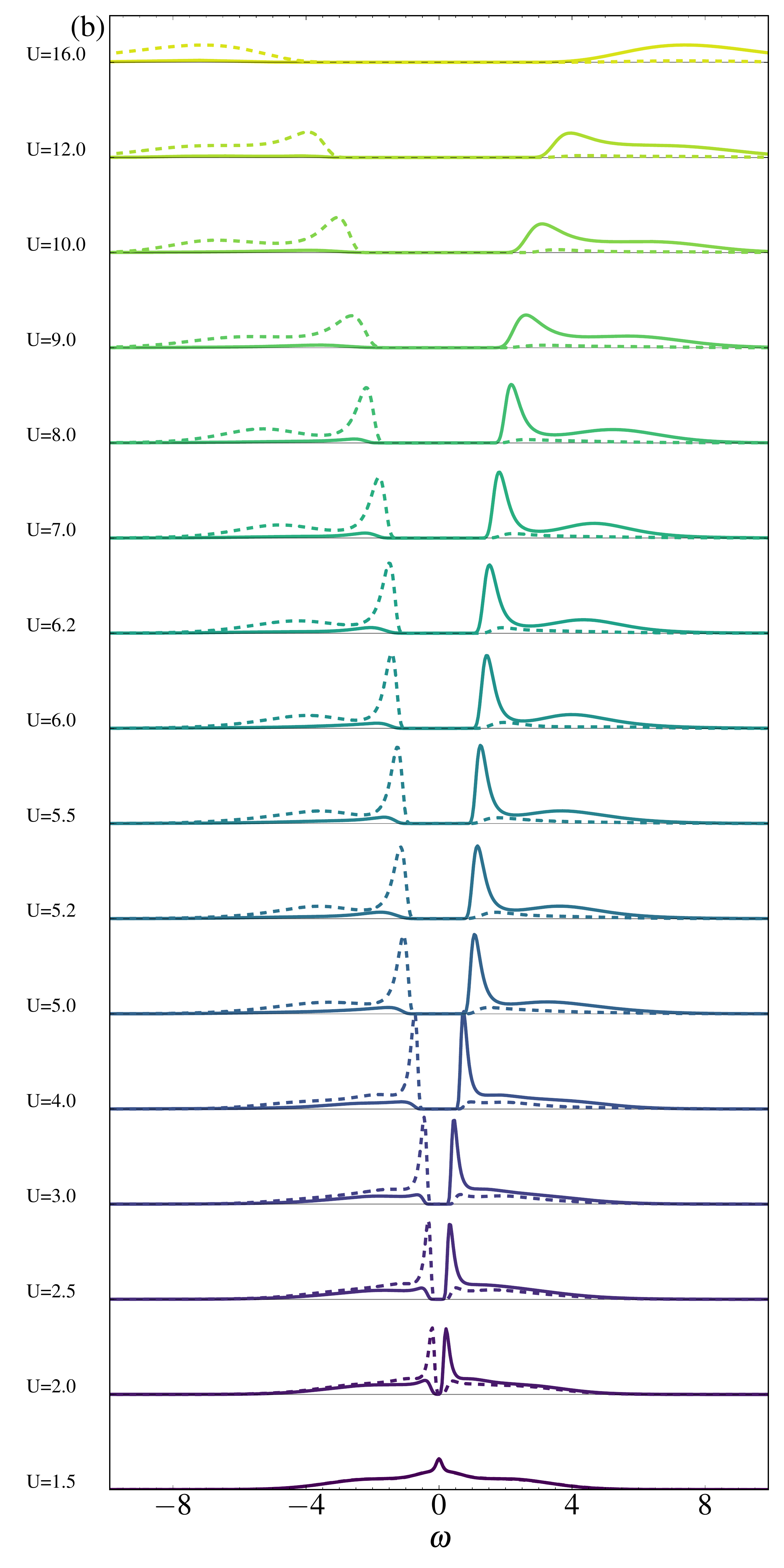}}
\caption{(a) $N(\omega)$ in the AF state along with (b) its two spin projections, $N_{\uparrow}(\omega)$ and $N_{\downarrow}(\omega)$, for $T=1/20$ and different values of $U$. Normal state solutions are shown with grey lines. This figure extends Fig.~\ref{fig3}a,b of the main text, where fewer values of $U$ are shown. 
}
\label{figS7}
\end{figure*}

\FloatBarrier


%

\end{document}